\documentclass[a4paper,superscriptaddress,showkeys,prX,showpacs,twocolumn]{revtex4-1}
\usepackage{amssymb, amsmath, amsfonts}
\usepackage{graphics}
\usepackage{pst-all}
\usepackage{hyperref}
\usepackage[displaymath,mathlines,running]{lineno}
\usepackage{relsize}
\usepackage[utf8]{inputenc}
\usepackage{IEEEtrantools}
\usepackage{epsfig}
\usepackage[toc]{appendix}
\newcommand{\ud}{\,\mathrm{d}}

\begin{document}
\title{Spin coherence on the ferromagnetic spherical surface}
\author{A. R. Moura}
\email{antoniormoura@ufv.br}
\affiliation{Departamento de F\'{i}sica, Universidade Federal de Vi\c{c}osa, 36570-900, Vi\c{c}osa, Minas Gerais, Brazil}
\date{\today}
\begin{abstract}

Spintronics on flat surfaces has been studied over the years, and the scenario is relatively well-known; however,
there is a lack of information when we consider non-flat surfaces. In this paper, we are concerned
about the spin dynamics of the ferromagnetic model on the spherical surface. We use the Schwinger bosonic
formalism for describing the thermodynamics of spin operators in terms of spinon operators. Opposite to the 
flat two-dimensional model, which is disordered at finite temperature, the curvature of the spherical surface
provides non-zero critical temperature for Schwinger boson condensation, which characterizes order at
finite temperature even in the absence of external magnetic fields. The thermodynamics is then analyzed
in the low-temperature regime. In addition, we consider the presence of both static and oscillating magnetic 
fields, the necessary condition for inducing the ferromagnetic resonance, and we show systematically that the 
studied model is well-described by $SU(2)$ coherent states, which provides the correct dynamics of the 
magnetization. The archived results can be applied for describing a diversity of experiments such as 
spin superfluidity, angular momentum injection by spin pumping and spin-transfer torque in non-conventional 
junctions, magnon dissipation, and magnetoelectronics on the spherical surface.

\end{abstract}
  
\pacs{}
\keywords{Ferromagnetic Ressonance; Spherical surface; Schwinger bosons}

\maketitle

\section{Introduction and motivation}

The continuous progress in spintronics has been motivated and supported by the potential realization
of technologies based on spin degrees of freedom in favor of the electrical ones.
Through a simple point-of-view, one of the principal purposes of spintronics is designing devices that work using
spin currents as a substitute for the (electrical) charge currents (for an extensive review of spintronics, see Ref. 
\cite{science294.1488} and \cite{rmp76.323}). Since spin currents occur in both normal metal 
and insulators, the applicability of spintronic devices is naturally higher than that one based on pure
electronic transport. Spin currents can arise due to the Spin Hall Effect (SHE) \cite{prl85.393,rmp87.1213}, 
the Spin Seebeck Effect (SSE) \cite{nature455.7214,natmat9.894,junsaku}, or through Spin Pumping 
(SP) from ferromagnetic resonance (FMR) \cite{prb66.224403,jap97.10c715,nature464.262,prb83.144402,prb89.174417}. 
On the other hand, the detection of spin current is obtained by converting it into a charge 
current through the Inverse Spin Hall Effect (ISHE) \cite{apl88.182509,prl98.156601,nature442.7099} 
or the Inverse Rashba-Edelstein Effect (IRRE) \cite{prl112.096601}. One can use the 
Spin-Transfer Torque (STT) experiment for verifying spin current transport \cite{prb66.014407} as well.   

In general, spintronic experiments involve flat surfaces and, therefore, there are no curvature effects 
in the thermodynamics of spin transport. However, the role of non-flat surfaces should be interesting
for non-conventional geometric devices. For example, medical researches have widely used hollow magnetic 
nanoparticles as drug transporter \cite{jnn12.2943,nanoscale19.9004,rsc9.25094}. At the same time, 
Hsu {\it et al.} showed the realization of the thin-film transistor on spherical surfaces 
\cite{apl81.1723,jap95.705}. From the theoretical point-of-view, spherical surfaces have been used for 
studying the role of curved two-dimensional space in phase transition such as Berezinskii-Kostertiz-Thouless 
(BKT) transition \cite{prl53.691,prd43.1314,jmmm513.167254} and Bose-Einstein Condensation (BEC)
\cite{apj87.924,prl123.160403}. For the latter case, many experiments of ultra-cold atoms on 
spherical bubbles have been proposed \cite{epl67.593,pra74.023616}; however, they require 
complex microgravity conditions to avoid the particles fall to the bottom of the trap \cite{science328.1540,prl123.240402,npjmg5.1}. 
Topological structures on curved manifold also were investigated in recent years. Kravchuck {\it et al.} studied 
out-of-surface vortices \cite{prb85.144433} and skyrmions \cite{prb94.144402} on spherical surfaces; curvature effects were
shown to be associated with effective magnetic interactions that provide the spin field on
curved manifolds \cite{jpamt48.125202,prl112.257203}; Sloika {\it et al.} determined the topological
structure of the magnetization on spherical shells in terms of geometrical parameters. A review of 
topological spin field excitations on curved spaces can be found in Ref. \cite{jpdap49.363001}.  

In this article, we use the Schwinger bosonic formalism for investigating the magnetization thermodynamics of
the ferromagnetic (FM) model on the spherical manifold. Despite the two-dimensional surface, the spherical
model presents some three-dimensional characteristics. Indeed, opposite to the flat two-dimensional
model, we find a finite phase transition temperature for all spin values. We choose the Schwinger 
Bosons Mean-Field Theory (SBMFT) because of its versatility for describing both ordered and disordered phases; however, in the
present article, we are mainly interested in the low-temperature regime. Although SBMFT can be improved by taking into account
Gaussian corrections in the mean-field parameters \cite{prl78.2216}, the mean-field fluctuations have been mostly applied
in frustrated antiferromagnetic (AFM) models \cite{prb96.174423,prb98.184403,prb100.104431}, whilst the usual SBMFT seems to
describe reasonably well the FM model, which is less susceptible to quantum fluctuations. In addition, 
we demonstrate that the Schwinger representation on the spherical surface does not present the pathological 
problem observed in flat space \cite{prb66.104417}. We show that the interaction with the oscillating magnetic 
field provides $SU(2)$ coherent states, which present similar points when compared with the $U(1)$ version. 
The magnetization and magnetic susceptibility are then determined using the $SU(2)$ coherent states of
the Schwinger bosons, and the results are in agreement with the expected ones.

\section{Model and Formalism}

We consider the ferromagnetic insulator described by the Hamiltonian $H(t)=H_0+V(t)$, where
the time-independent part is given by
\begin{equation}
\label{eq.H0}
H_0=-J\sum_{\langle ij\rangle}\vec{S}_i\cdot\vec{S}_j-g\mu_B B^z\sum_i S_i^z,
\end{equation}
in which the sum is taken over nearest-neighbor spins on the spherical surface, and
$J>0$ is the exchange coupling. The time-dependent term represents the interaction with the oscillating 
magnetic field, being expressed by
\begin{equation}
\label{eq.interaction}
V(t)=-g\mu_B B^x(t)\sum_i S_i^x.
\end{equation}
We are adopting both magnetic fields $B^x$ and $B^z$ as uniform fields based on the reduced dimensions
of the samples in spintronic experiments. The interaction $V(t)$, which is treated according to the interaction
picture, provides the coherent states necessary for the description of the magnetization precession.

\begin{figure}[h]
\centering \epsfig{file=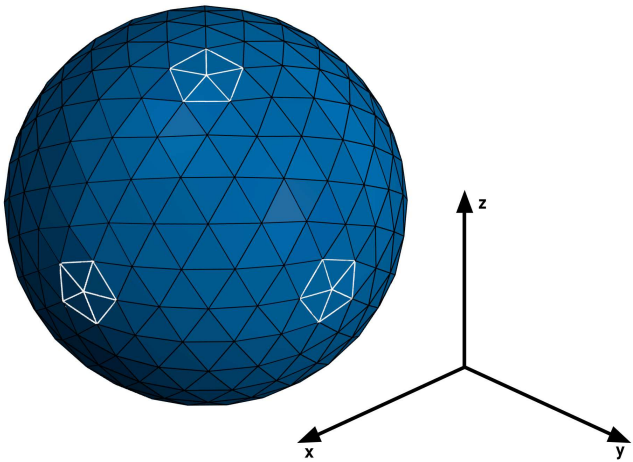,width=1.0\linewidth}
\caption{Spherical tessellation of the icosahedron. The vertices of the polyhedron provide sites with 
five neighbors, while the other sites have six ones.}
\label{fig.icosahedron}
\end{figure}

Opposite to the planar square lattice, in which each site always has four neighbors, it is impossible to 
build a regular discrete lattice on a spherical surface due to its topology. For a review of the lattice representations on the sphere, see Ref. \cite{drna9.1}. For avoiding the singularities at the poles of 
the geographic coordinates grid, we consider geometric tessellations based on the icosahedron \cite{cg32.1442}. 
Each side of the icosahedron is subdivided in a regular lattice and then, the sites are then projected onto the
spherical surface. As one can see in fig. \ref{fig.icosahedron}, most sites have six neighbors, while the 
vertices sites have five neighbors. The exact grid adopted is not so relevant since we use a continuous 
representation of the Hamiltonian (\ref{eq.hamiltonian}); however, the number of neighbors $z$ 
(coordination number) is important, and we will consider $z=6$. Note that due to the triangular symmetry
of the lattice, the antiferromagnetic model on the spherical surface will be frustrated, which requires
special treatment for decoupling the quartic terms \cite{prb49.1131}. In addition, frustrated models are
sensible to quantum fluctuations at low-temperatures and, in this case, the SBMFT need to be endowed with
Gaussian corrections in the mean-field parameters \cite{prb96.174423,prb98.184403,prb100.104431}. Curiously, the spherical
curvature implies changes in the winding number of topological solutions. The uniform solution, which 
presents the spin field align to a fixed direction, has winding number $Q=1$, while $Q=0$ for the 
lowest-energy skyrmion solution ($Q=0$ and $Q\neq 0$ for the ground-state and skyrmion solutions, respectively, 
when we consider the flat two-dimensional space). In addition, the correct development of skyrmion-kind 
excitations requires the uniaxial anisotropy $(\vec{S}\cdot\vec{n})^2$, where $\vec{n}$ is 
the outward normal vector, as pointed by Kravchuk {\it et al.} \cite{prb94.144402}. Here, since
we do not interested in topological solutions, and due to the magnetic field $B^z$, which aligns
the spin field along the z-axis, we do not consider the uniaxial anisotropy. 

At low-temperature, spin operators are usually treated by using the Holstein-Primakoff (HP) 
bosonic representation\cite{pr58.1098};  however, HP bosons are inaccurate for representing disordered magnetic 
phases. The more appropriate representation is obtained through Schwinger bosons, which apply 
to both ordered and disordered phases \cite{prb38.316,prb40.5028}. 
The spin operators are then replaced by two kinds of bosonic operators and written as 
$S_i^+=a_i^\dagger b_i$, $S_i^-=b_i^\dagger a_i$, and $S_i^z=(a_i^\dagger a_i-b_i^\dagger b_i)/2$, 
where $a_i^\dagger$ ($b_i^\dagger$) creates a spinon with spin $1/2$ (-$1/2$) in the site $i$. 
For ensuring the commutation relation $[S_i^a,S_j^b]=i\delta_{ij}\epsilon_{abc}S_i^c$, it is 
necessary to fix the number of bosons on each site through the local constraint 
$a_i^\dagger a_i+b_i^\dagger b_i=2S$. Note that the spin operators are invariant under the $U(1)$ gauge 
transformation $a_i\to e^{i\psi}a_i$ and $b_i\to e^{i\psi}b_i$, where $\psi$ is a global phase.
Therefore, the Hamiltonian $H_0$ is written as

\begin{IEEEeqnarray}{rCl}
\label{eq.hamiltonian}
H_0&=&-\frac{J}{2}\sum_{\langle ij\rangle} (:\mathcal{F}_{ij}^\dagger \mathcal{F}_{ij}:-2S^2)+\sum_i \lambda_i(\mathcal{F}_{ii}-2S)-\nonumber\\
&&-\frac{g\mu_B B^z}{2}\sum_i(a_i^\dagger a_i-b_i^\dagger b_i)
\end{IEEEeqnarray}
in which we defined the bond operator 
$\mathcal{F}_{ij}=a_i^\dagger a_j+b_i^\dagger b_j$ and $:\ :$ represents the normal ordering operator. 
The constraint is implemented by a local Lagrange multiplier $\lambda_i$, and the quartic order term 
is decoupled by introducing the auxiliary field $F_{ij}=\langle\mathcal{F}_{ij}\rangle$ through 
the Hubbard-Stratonovich transform 
$\mathcal{F}_{ij}^\dagger \mathcal{F}_{ij}\to F_{ij}(\mathcal{F}_{ij}^\dagger+\mathcal{F}_{ij})-F_{ij}^2$.
As usual, we adopt the mean-field theory and $F_{ij}$ is replaced by the uniform field 
$F$. We also approximate the Lagrange multiplier by a uniform parameter $\lambda$ that 
implies boson conservation only on average. In general, SBMFT is successful to describe the thermodynamic
of magnetic models but is also known that SBMFT gives the incorrect local spin-spin correlation due to the 
missing of a multiplicative $2/3$ factor \cite{prb38.316,auerbach}. The problems with the mean-field theory are more evident 
when quantum fluctuations are more relevant, as occurs in frustrated AFM models, for example. 
To correct this inconvenience, Gaussian corrections can be implemented in the SBMFT \cite{prl78.2216}; 
however, since we are considering the FM model in the coherent state, quantum fluctuations have a minor effect 
(opposite to frustrated AFM models) and the mean-field theory is supposed to provide reasonable results.

The diagonalization of Eq. (\ref{eq.hamiltonian}) involves the continuous limit, which also
provides an easy method for including the curvature effect. In the long-wavelength limit, we adopt
the second-order expansion $a_j\approx a_i+\varepsilon^k \partial_k a_i+(\varepsilon^k\varepsilon^l/2) \partial_k\partial_l a_i$, where $\varepsilon^k$ are infinitesimal displacements along the polar or
azimuth directions, and the sums in $k$ and $l$ are implicit. The expansion results in 
$\sum_{\langle ij\rangle} a_i^\dagger a_j\approx \int\ud\Omega z\sigma [a^\dagger a+(\varepsilon^2/4)a^\dagger\nabla^2a]$, where $d\Omega=\sin\theta d\theta d\varphi$ with $0\leq\theta\leq\pi$
and $0\leq\varphi<2\pi$ being the polar and azimuth angles, respectively. The (angular) density of states 
on the spherical surface is defined as $\sigma=\mathcal{N}/4\pi$, and we choose $\varepsilon$ in order to 
obtain $R^2=\sigma\varepsilon^2$, where $R$ is the sphere radius. We write the continuous $a(\theta,\varphi)$
operator as the following spherical harmonic expansion
\begin{equation}
\label{eq.spherical_exp}
a(\theta,\varphi)=\frac{1}{\sqrt{\sigma}}\sum_L a_L Y_L(\theta,\varphi),
\end{equation}
where $L$ stands for a compact notation of $lm$, with $l$ integer and $m=-l,-l+1,\ldots,l-1,l$.
The same procedure is applied for the $b$ operator. The $\sigma$ factor is included for ensuring the same 
number of bosonic modes on both bases, {\it i.e.} $\sum_i a_i^\dagger a_i=\sum_La_L^\dagger a_L$. 
In addition, we introduce a superior limit $l_\textrm{max}$ for the $l$ sum. This restriction is 
expected since in the continuous representation we have considered smooth operators in the expansion of $a_j$, 
and spherical harmonics with fast oscillations ($l>l_\textrm{max}$) have negligible contributions. 
Therefore, assuming the lattice parameter as a space cutoff, and 
adopting one-to-one correspondence between the representations, we obtain 
$\mathcal{N}=\sum_i=\sum_L=(l_\textrm{max}+1)^2$. Using the properties of the spherical harmonics,
it is straightforward to obtain
\begin{equation}
\label{eq.hamiltonian_diag}
H_0=E_0+\sum_L \left(\epsilon_L^{(a)} a_L^\dagger a_L+\epsilon_L^{(b)} b_L^\dagger b_L\right),
\end{equation}
where $E_0/\mathcal{N}=zJF^2/2-2S(zJF-\mu)$ is a constant energy, $\epsilon_L^{(a)}=\epsilon_L-\mu-g\mu_B B^z/2$,
$\epsilon_L^{(b)}=\epsilon_L-\mu+g\mu_B B^z/2$, $\mu=zJF-\lambda$ plays the rule of a chemical potential
(we can also define $-\mu$ as the gap energy), and $\epsilon_L=zJFl(l+1)/4\sigma$ is the $(2l+1)$-fold 
degenerate energy mode. In Appendix (\ref{appendix_LSW}), we developed the Linear Spin-Wave (LSW) theory 
using the HP bosonic formalism. At very low temperatures, the results obtained from both formalism are
identical.

\subsection{Generation of the coherent states}
For obtaining the magnetization precession, it is necessary to apply an oscillating magnetic field
perpendicular to the static field $B^z$. The relation between the coherent states and the oscillating field is
deduced as follows. Through the spherical Schwinger bosons, we write $\sum_i S_i^x=(J^+ + J^-)/2$, 
where we define the many-site operators $J^+=J^x+i J^y=\sum_L a_L^\dagger b_L$ ($=\sum_i a_i^\dagger b_i$)
and $J^-=(J^+)^\dagger$, which follow the Lie algebra
\begin{equation}
[J^a,J^b]=i\epsilon_{abc} J^c,
\end{equation} 
in which the structure constants are given by the Levi-Civita symbol $\epsilon_{abc}$. 
Therefore, the Hamiltonian $H$ is expressed in terms of the generators of the group $SU(2)$. We choose a basis
in terms of the eigenstates of the operator $J^z=\sum_L (a_L^\dagger a_L - b_L^\dagger b_L)/2$, 
namely $|j j_z\rangle$, with $j$ integer or half-integer and $j_z=-j,-j+1,\ldots,j-1,j$. The Casimir operator 
$J^2=(J^z)^2+(J^+J^-+J^-J^+)/2$ satisfies $[J^z,J^2]=0$ as well as the eigenvalue equation 
$J^2|j j_z\rangle=\hbar j(j+1)|j j_z\rangle$, as observed for standard angular momentum operators. In our 
case, $J$ represents the angular momentum sum of $\mathcal{N}$ sites with spin $S$ on the spherical surface, 
and the Hilbert space is spanned as the direct sum of irreducible representations according to $j$. Since
$\mathcal{N}$ is odd, the minimum $j_\textrm{min}=S$ is obtained when $\mathcal{N}-1$ sites are paired with
opposite spin, and the maximum $j_\textrm{max}=\mathcal{N}S$ occurs when the $\mathcal{N}$ sites are unpaired.
Here and henceforth, we will consider a fixed $j$ and define the lowest-weight (extremal state) $|j -j\rangle$
as $|\Psi_0\rangle$. 

Any element $g\in SU(2)$ can be expressed as 
\begin{equation}
g=D(\theta,\varphi) h,
\end{equation}  
where $h=\exp(i\alpha J^z)$ is an element of the group $U(1)$, with $\alpha$ being a real parameter, 
and $D(\theta,\varphi)$ is the coset representative of $SU(2)/U(1)$, {\it i.e.} the two-dimensional sphere $S^2$. 
One can show that $D(\theta,\varphi)=\exp(i \theta\vec{m}\cdot\vec{J})$, 
with $\vec{m}=(\sin\varphi,-\cos\varphi,0)$ and, therefore, $D(\theta,\varphi)$ provides (expect for a 
multiplicative constant chosen to be unity) the coherent state $|\vec{n}\rangle$ 
given by \cite{rmp62.867,perelomov}
\begin{equation}
|\vec{n}\rangle=D(\theta,\varphi)|\Psi_0\rangle.
\end{equation}
Physically, the state $|\vec{n}\rangle$ represents the classical vector 
$\vec{n}=(\sin\theta\cos\varphi,\sin\theta\sin\varphi,\cos\theta)$, and $D(\theta,\varphi)$ is the 
so-called generalized displacement operator. For our present purpose, it is
more convenient to express the operator $D$ as
\begin{equation}
D(\zeta)=e^{\zeta J^+-\bar{\zeta} J^-},
\end{equation}
in which $\zeta=-(\theta/2)\exp(-i\varphi)$ is a complex parameter. Note that all spins are described
by the same angles $\theta$ and $\varphi$, which is characteristic of magnetization precession. 
In addition, the above expression makes clear the similarity between $D(\zeta)$ and the displacement 
operator of the $U(1)$ coherent state theory \cite{pr131.2766}, which justifies the chosen name. 

Returning to the Hamiltonian $H$, we write the time-dependent average of an operator $A(t)$
in the Interaction picture
\begin{equation}
\langle A(t)\rangle=\langle\Psi_0| U^\dagger(t,-\infty)\hat{A}(t)U(t,-\infty)|\Psi_0\rangle,
\end{equation}
with the caret denoting time evolution according $H_0$ and
\begin{equation}
U(t,-\infty)=T_t\exp\left(-\frac{i}{\hbar}\int_{-\infty}^t \hat{V}(t^\prime)dt^\prime\right),
\end{equation}
where $T_t$ is the time-ordering operator. Here, we consider an adiabatic process from the dim 
past ($t\to\infty$), for which the interaction is off and the ground-state is $|\Psi_0\rangle$, 
to the present time with full Hamiltonian $H(t)=H_0+V(t)$. It is a straightforward procedure
to show that the argument of the exponential of $U$ is given by $\zeta(t)J^+-\bar{\zeta}(t)J^-$, where
\begin{equation}
\zeta(t)=\frac{ig\mu_B}{2\hbar}\int_{-\infty}^t B^x(t^\prime)\exp\left[\frac{i}{\hbar}(\epsilon_0^{(a)}-\epsilon_0^{(b)})t^\prime\right].
\end{equation}
Despite an irrelevant phase factor, we can show that
\begin{equation}
U(t,-\infty)|\Psi_0\rangle=D(\zeta)|\Psi_0\rangle=|\zeta\rangle,
\end{equation}
and the oscillating magnetic field generates the $SU(2)$ coherent state $|\zeta\rangle$.
As usual in experiments, we adopt a monochromatic magnetic field with frequency $\omega_\textrm{rf}$, 
for which we obtain
\begin{equation}
\label{eq.zeta}
\zeta(t)= \frac{\gamma B^x}{2(\gamma B^z-\omega_\textrm{rf}-i\eta)}e^{i(\gamma B^z-\omega_\textrm{rf})t},
\end{equation} 
where $\gamma=g\mu_B/\hbar$ is the gyromagnetic ratio, and the factor $\eta$ is added for ensuring
the convergence in the limit $t\to-\infty$. 

\section{Thermodynamics at low-temperature}
Since the Hamiltonian $H_0$ is described by coherent states, we can write the partition function
as $Z_0=\int D[\bar{a},a]D[\bar{b},b]\exp[-\mathcal{S}/\hbar]$, where the integration measures 
represent the integration on each point on the continuous spherical surface (for more details, see 
Appendix \ref{appendix_su2cs}). Note that, since we have adopted the 
mean-field replacement for $F$ and $\lambda$, the partition function does not involve path integration 
over $F$ and $\lambda$. In the spherical harmonic space, the action is given by

\begin{IEEEeqnarray}{rCl}
\mathcal{S}&=&\beta\hbar E_0+\frac{\beta\hbar}{2}\sum_{i\omega_p}\sum_L\left[\bar{a}_L(-i\hbar\omega_p +\epsilon_L^{(a)})a_L+\right.\nonumber\\ 
&&\left.+\bar{b}_L(-i\hbar\omega_p +\epsilon_L^{(b)}) b_L+\textrm{h.c.}\right],
\end{IEEEeqnarray}
with the Matsubara frequencies $\omega_p=2\pi p/\beta\hbar$, $p\in\mathbb{Z}$. Here, for convenience, we use 
the same notation $a_L$ and $b_L$ for representing the fields associated with the correspondent 
operators $a_L$ and $b_L$, respectively. The meaning of $a_L$ and $b_L$ should be clear from the context but, 
if necessary, a comment will be inserted. After integrating out the fields, we obtain the mean-field free energy
\begin{equation}
F_\textrm{MF}=E_0+\frac{1}{2\beta}\sum_{i\omega_p}\sum_L\ln(\beta M)
\end{equation}
where $M=\epsilon_L^{(a)}(I+\sigma_z)\otimes I/2+\epsilon_L^{(b)}(I-\sigma_z)\otimes I/2-i\hbar\omega_p I\otimes\sigma_z$. Provided that $F_\textrm{MF}=E_0$ in the limit $T\to 0$, the Matsubara frequency sum
results in
\begin{equation}
F_\textrm{MF}=E_0+\frac{1}{\beta}\sum_L \ln\left[(1-e^{-\beta\epsilon_L^{(a)}})(1-e^{-\beta\epsilon_L^{(b)}})\right].
\end{equation}

In the presence of the magnetic field $B^z$, a broken symmetry takes place at low-temperatures, 
as expected. The more interesting case occurs in the limit of vanishing magnetic field $B^z$. For
a flat two-dimensional model (with only short-range interactions and continuous symmetry) 
free of magnetic fields, there is no ordering at finite temperature; however,
the situation is different for the spherical manifold. Indeed, the impossibility of magnetic ordering at
finite temperature comes from the Mermin-Wagner theorem, which requires thermodynamic limit for
yielding low-energy Goldstone modes. Insofar as the spherical surface is compact, the Mermin-Wagner theorem
is not applicable and would be possible to observe spontaneously broken symmetry at finite temperatures.

In the Schwinger bosonic formalism, an ordered state is related to the BEC of the spinon modes \cite{prb40.5028}. 
Below a critical temperature $T_c$, the chemical potential vanishes, and the bosons condensate in 
the minimal energy state $\epsilon_{l=0}$, while for $T>T_c$, $\mu<0$. The connection between the magnon-picture 
and the spinon-picture is given as following. The up-spin spinon band represents the ordered ground state, while the down-spin 
spinon is associated with the spin excitation. Therefore, a magnon is described by an up spinon that suffers a 
transition to a down-spin spinon. Note that creation/annihilation spinon processes only happen in pairs and
free spinons are not observed in FM models. For taking into account the BEC, we separate the $l=0$ term from the sum, which is 
indicated by a prime, and write the free energy per site as

\begin{IEEEeqnarray}{rCl}
\label{eq.FMF}
\frac{F_\textrm{MF}}{\mathcal{N}}&=&f_\textrm{MF}=\frac{\xi^2}{2zJ}-2S\xi[1+(1-\rho)\Delta]+\nonumber\\
&&+\frac{2}{\beta\mathcal{N}}{\sum_L}^\prime\ln\left[1-e^{-\frac{1}{t}\left(\frac{l(l+1)}{4\sigma}+\Delta\right)}\right],
\end{IEEEeqnarray}
where $\xi=zJF$ is the energy scale, $\Delta=-\mu/\xi$ is the dimensionless gap energy, $t=k_BT/\xi$ is the reduced
temperature, and $\rho=\mathcal{N}_0/\mathcal{N}$ measures the condensation level. At low-temperatures,
the exponential decreases fast enough, and we can evaluate the sum replacing it by an integral. Using
the Mercator series $\ln(1-x)=-\sum_{k=1}^\infty x^k/k$, we obtain
\begin{equation}
f_\textrm{MF}=\frac{\xi^2}{2zJ}-2S\xi[1+(1-\rho)\Delta]-\frac{2\xi t^2}{\pi}\textrm{Li}_2(g e^{-1/2t\sigma}),
\end{equation}
where $\textrm{Li}_s$ is the polylogarithmic function of order $s$ and $g=\exp(-\Delta/t)$ is the fugacity.
The parameters $\xi$ and $\Delta$ are then determined by the extremum conditions $\partial f_\textrm{MF}/\partial \Delta=0$ and $\partial f_\textrm{MF}/\partial \xi=0$, which provide
\begin{equation}
\label{eq.rho}
\rho=1-\frac{t}{\pi S}\textrm{Li}_1(ge^{-1/2t\sigma}),
\end{equation}
and
\begin{equation}
\label{eq.xi}
\frac{\xi}{zJ}=2S-\frac{2t^2}{\pi}\textrm{Li}_2(ge^{-1/2t\sigma})
\end{equation}
respectively. Observe that $\Delta=0$ and $0<\rho\leq1$ below the critical temperature, while $\Delta>0$
and $\rho=0$ for $T>T_c$. The reduced critical temperature $t_c$ is then obtained making
$\rho=0$ and $\Delta=0$ in Eq. (\ref{eq.rho}), which results in the self-consistent equation
\begin{equation}
\label{eq.tc}
t_c=\frac{\pi S}{\textrm{Li}_1(e^{-1/2t_c\sigma})}
\end{equation}

\begin{figure}[h]
\centering \epsfig{file=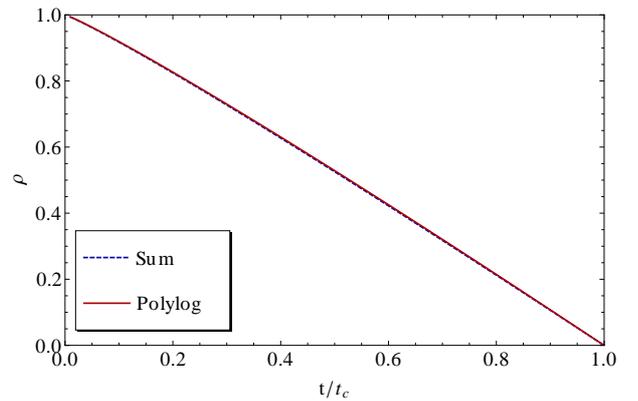,width=1.0\linewidth}
\caption{The condensation $\rho$ as a function of the reduced temperature. The results obtained 
by using the sum on $l$ and $m$ and from the continuum (polylogarithmic) approximation are close. Here, we 
consider $\mathcal{N}=10^6$, $S=1$, $t_c=0.26$ (Polylog result) and $t_c=0.25$ (Sum result).}
\label{fig.rho_t}
\end{figure}

\begin{figure}[h]
\centering \epsfig{file=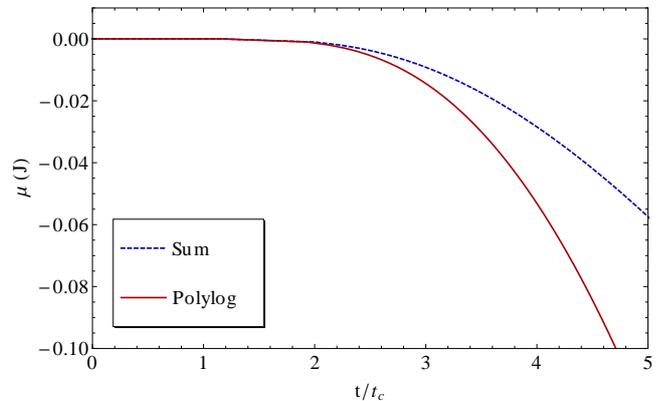,width=1.0\linewidth}
\caption{The chemical potential determined through the polylogarithmic approximation and by
the sum over $L$. Here, we adopt $\mathcal{N}=10^6$, $S=1$ and $t_c=0.26$.}
\label{fig.mu_t}
\end{figure}

Since both $\rho$ and $\xi$ parameters depend only on the reduced temperature, it
is easy to solve the equations. One can also obtain the above equations without performing
the continuum approximation. Figure \ref{fig.rho_t} shows the results for the condensation 
by using the two approaches, directly from the sum and through the polylogarithmic equation. As one can see,
both curves are close at low-temperatures. For $S=1$ and $\mathcal{N}=10^6$, we obtain 
$t_c=0.25$ evaluating the sum over $L$, and $t_c=0.26$ when we use Eq. (\ref{eq.tc}). The chemical 
potential is also determined using both methods and the results are shown in fig. \ref{fig.mu_t}. 
Again, the difference between the polylogarithmic and the result obtained from the sum is small 
at low-temperatures. 

\begin{figure}[h]
\centering \epsfig{file=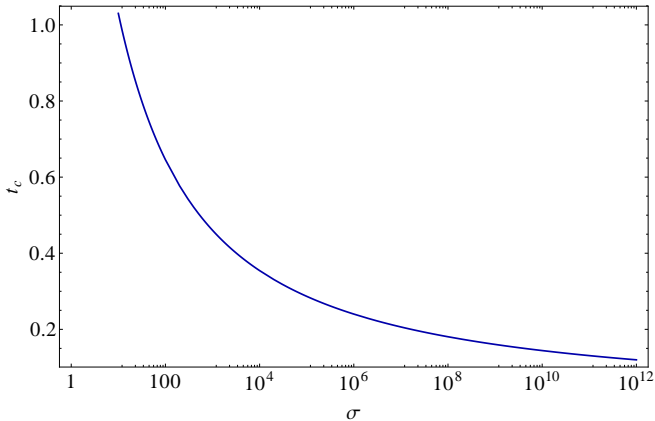,width=1.0\linewidth}
\caption{The reduced critical temperature dependence on the density of sites. Locally, the 
limit $\sigma\to\infty$ reflects the flat space, which provides $t_c\to0$.}
\label{fig.tc_sigma}
\end{figure}

The reduced critical temperature is shown in fig. \ref{fig.tc_sigma}. The critical
temperature is a decreasing function of increasing density of states, and $t_c$ tends to zero in 
the limit of $\sigma\to\infty$. Indeed, when $\sigma$ is very large, the distance between 
nearest-neighbors on the surface is small and the curvature effect is negligible. Locally, the 
short-range interaction resembles the flat-space interaction, and we recovery the flat two-dimensional
result, for which the critical temperature $T_c=0$. In addition, there is no critical spin 
and the transition temperature is finite for any spin value, including the classical limit $S\gg 1$. Here, for
sake of simplicity, we consider $S=1$.

\begin{figure}[h]
\centering \epsfig{file=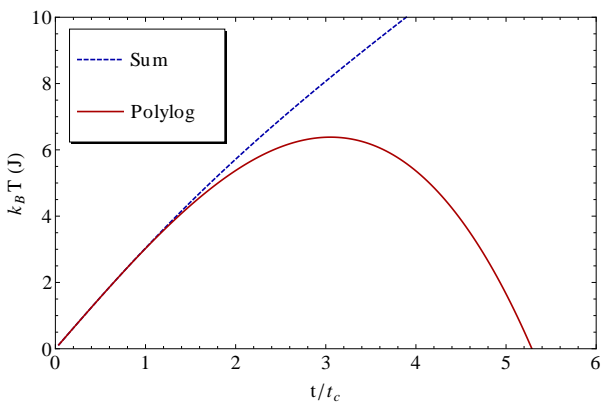,width=1.0\linewidth}
\caption{The relation between the physical $T$ and the reduced $t$ temperatures. For $t>t_c$, the 
approximated polylogarithmic equations present spurious results.}
\label{fig.T_t}
\end{figure}

\begin{figure}[h]
\centering \epsfig{file=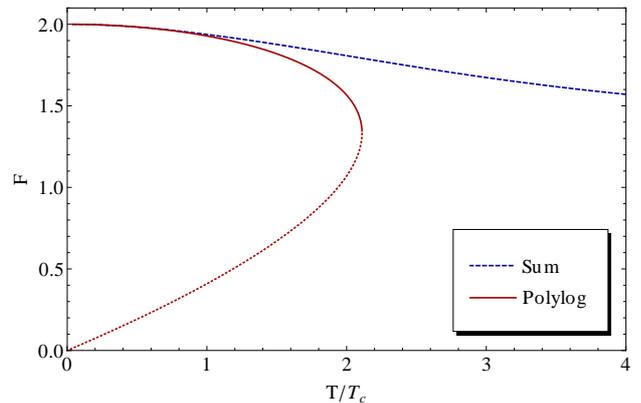,width=1.0\linewidth}
\caption{The average ferromagnetic bond $F$ as function of $T$. For $t>t_c$, we obtain the dotted branch, 
which is non-physical. Here, $T_c=t_c\xi(t_c)$, with $t_c=0.26$.}
\label{fig.F_T}
\end{figure}

The polylogarithmic approximation is very reasonable in the ordered state; however, when $t>t_c$ the
approximated equations are not accurate, and any result should be determined by using the equations
written in terms of the sum over $L$. The problem occurs due to the slow decreasing of the exponential in Eq. (\ref{eq.FMF})
when $k_B T>zJF$, which precludes the series expansion of the logarithmic function. To verify the spurious result
of the approximation at high temperature, we analyze the relation between the physical 
and reduced temperature given by $T(t)=t\xi(t)$ and shown in fig. \ref{fig.T_t}. The function
$T(t)$ obtained by the approximation shows a maximum at $t_\textrm{max}\approx 3.1 t_c$, 
and for $t>t_\textrm{max}$, $T$ decreases with increasing $t$, for which we obtain a non-physical result. 
Indeed, because of the decreasing behavior of $T(t)$, $F(T)$ shows is an increasing function of increasing 
$T$, as shown in fig. \ref{fig.F_T}, where the dotted branch represents the spurious result obtained 
through the approximation for $t>t_c$. The correct behavior, which shows a decreasing 
$F$ for increasing $T$, is obtained when we use the equations for $\xi$ and $\rho$ written using the
sum over $L$. The spurious branch also appears in the three-dimensional ferromagnetic model, 
as pointed in Ref. (\cite{prb66.104417}) (in this case, the Schwinger formalism presents serious problems
for describing spin models with $S\gtrsim 0.15$). Since we are interested in the ordered 
state ($t<t_c$), we keep the polylogarithmic approximation. 

It is important to note that the limit of high temperatures ($t\gg t_c$) can not be described
by the equations developed above. Provided that we consider only long-wavelength spin-wave in
the continuous description of the Hamiltonian, high-energy excitations are not covered by the
present model. For including the high-energy spectrum, one should write 
$\sum_{\langle ij\rangle}a_i^\dagger a_j=\sum_L \gamma_L a_L^\dagger a_L$, where the structure
factor is given by
\begin{equation}
\gamma_L=\sum_{\eta_\theta,\eta_\varphi}\int\ud\Omega\bar{Y}_L(\theta,\varphi)Y_L(\theta+\eta_\theta,\varphi+\eta_\varphi),
\end{equation}
and $\eta_{(\theta,\varphi)}$ are the angular nearest-neighbor positions. Expanding the
neighbor sites around $(\theta,\varphi)$, we obtain $\gamma_L\approx z-zl(l+1)/4\sigma$, which
recoveries the Hamiltonian (\ref{eq.hamiltonian_diag}). Since we are interested in the low-energy
limit, it is not necessary consider the full structure factor.

When the magnetic field $B^z$ is included, the equations are obtained following the same steps, and the
only difference occurs in the condensation. In this case, since the $b$ modes acquired a gap due to the
Zeeman energy, only the $a$ bosons condensate in the $l=0$ state. The critical temperature, for example,
is then given by
\begin{equation}
t_c=\frac{2\pi S}{\textrm{Li}_1(e^{-1/2t_c\sigma})+\textrm{Li}_1(e^{-(1/2\sigma+\hbar\gamma B^z)/t_c})},
\end{equation}  
which recoveries Eq. (\ref{eq.tc}) when $B^z=0$.

\section{Magnetization dynamics}
\label{sec.mag_dynamics}
To determine the dynamics of the spin $\vec{S}_i(t)$, and so the surface magnetization
defined by $\vec{M}=(\gamma\hbar/\mathcal{N}\varepsilon^2)\sum_i\langle\vec{S}_i\rangle$, 
let us define the coherent state $|\zeta_i\rangle$.
Using the Baker-Campbell-Hausdorff (BCH) formula, the generalized displacement operator 
for the $i$ site is expressed as 
\begin{equation}
D(\zeta_i)=e^{\kappa_i J_i^+}e^{\ln(1+|\kappa_i|^2)J_i^z}e^{-\bar{\kappa_i}J_i^-},
\end{equation}
with $J_i^+=a_i^\dagger b_i$, $J_i^z=(a_i^\dagger a_i-b_i^\dagger b_i)/2$, and $\kappa_i=\exp(-i\varphi_i)\tan(\theta_i/2)$. 
When applied on the extremum state $|\psi_0\rangle$, for which
$S^z|\psi_0\rangle=-S|\psi_0\rangle$, the displacement operator provides
\begin{equation}
D(\zeta_i)|\psi_0\rangle=\frac{1}{(1+|\kappa_i|^2)^S}e^{\kappa_i J_i^+}|\psi_0\rangle.
\end{equation}
Since $|\psi_0\rangle$ represents the state with $n_a=0$ and $n_b=2S$, the operator $J_i^+$ can
be applied a maximum of $2S$ times, and the k-th operation results in
$(J_i^+)^k|\psi_o\rangle=[(2S)!k!/(2S-k)!]^{1/2}|n_a=S,n_b=2S-k\rangle$. Therefore, we obtain
\begin{equation}
|\zeta_i\rangle_{2S}=D(\zeta_i)|\psi_0\rangle=\sqrt{(2S)!}\sum_{n_a,n_b}\frac{u_i^{n_a}v_i^{n_b}}{\sqrt{n_a!n_b!}}|n_a,n_b\rangle,
\end{equation}
where the implicit local constraint $n_a+n_b=2S$ is assumed in the sum, which is indicated
by the subscript $2S$ in $|\zeta_i\rangle$, and we define the local parameters
\begin{equation}
u_i=\cos\left(\frac{\theta_i}{2}\right),
\end{equation}
and
\begin{equation}
v_i=\sin\left(\frac{\theta_i}{2}\right)e^{-i\varphi_i}.
\end{equation}
Due to the finite sum, the $SU(2)$ coherent states
are not eigenstates of the annihilation operators, as seen in the $U(1)$ representation; 
however, a straightforward procedure shows that
\begin{equation}
\label{eq.azeta}
a_i|\zeta_i\rangle_{2S}=\sqrt{2S}u_i|\zeta_i\rangle_{2S-1},
\end{equation}
and
\begin{equation}
\label{eq.bzeta}
b_i|\zeta_i\rangle_{2S}=\sqrt{2S}v_i|\zeta_i\rangle_{2S-1}.
\end{equation}
Therefore, $|\zeta_i\rangle$ plays a similar rule to that present in the $U(1)$
coherent state formalism and using the above equations, it is easy to show that 
the $SU(2)$ coherent states give the average $\langle a_i^\dagger a_i+b_i^\dagger b_i\rangle=2S$.

Considering $|\zeta|<\pi/2$, equation (\ref{eq.zeta}) gives the global angles
$\theta=\gamma B^x/(\gamma B^z-\omega_\textrm{rf}-i\eta)$ [for $|\zeta|>\pi/2$, we define 
the polar angle by $4\theta=\gamma B^x/(\gamma B^z-\omega_\textrm{rf}-i\eta)\ \textrm{mod}\ 2\pi$] 
and $\varphi=(\omega_\textrm{rf}-\gamma B^z)t$. The z-component of the magnetization is 
time-independent and given by $M^z=\gamma\hbar\langle\cos\theta\rangle\approx \gamma\hbar$ for small 
polar angles, as expected when $\vec{M}$ involves according to precession dynamics. For analyzing the 
perpendicular magnetization component, we define the complex field 
$M^\perp=M^x+i M^y=(\gamma\hbar/\mathcal{N})\sum_i a_i^\dagger b_i$, which yields the uniform magnetization
$M^\perp(t)=\gamma\hbar S e^{-i\gamma B^z t}\langle\sin\theta e^{-i\varphi}\rangle$, and for 
small $\theta$, we get
\begin{equation}
\label{eq.magnetization}
M^\perp(t)=\frac{M_s \gamma B^\perp(t)}{\gamma B^z-\omega_\textrm{rf}-i\eta},
\end{equation}
where $B^\perp(t)=\mu_0[H^\perp(t)+M^\perp(t)]=B^x e^{-i\omega_\textrm{rf} t}$ is a monochromatic field, 
and $M_s=\gamma\hbar S/\varepsilon^2$ is the saturation magnetization on the spherical surface. In the spherical harmonic-frequency space, 
the magnetization is written as $M_L^\perp(\omega)=\sum_{L^\prime}\chi_{LL^\prime}(\omega)H_{L^\prime}^\perp(\omega)$.
Note that for uniform monochromatic field and magnetization, the only accessible mode of the susceptibility 
is $\chi_{00}(\omega_\textrm{rf})$. Then, using Eq. (\ref{eq.magnetization}), we obtain the long-wavelength magnetic susceptibility 
$\chi_{00}=\chi^\prime+i\chi^{\prime\prime}$, where the uniform real and complex parts are given by
\begin{equation}
\chi^\prime=\frac{\omega_M(\omega_0-\omega_\textrm{rf})}{(\omega_0-\omega_\textrm{rf})^2+\eta^2}
\end{equation}
and
\begin{equation}
\chi^{\prime\prime}=\frac{\omega_M\eta}{(\omega_0-\omega_\textrm{rf})^2+\eta^2},
\end{equation}
respectively. In above equations, $\omega_0=\gamma\mu_0 H^z$ is the frequency of the lowest-energy magnons,
and $\eta$ is related to the Gilbert damping. As one can see, the magnetic response is maximum when magnetic fields 
satisfy $\gamma\mu_0 H^z=\omega_\textrm{rf}$, the resonating condition. 
Typical experiments involve resonating frequencies of the order of GHz and $H^z$ of the order of 
$10^{-1}$ T \cite{jap97.10c715,prl113.266602,prb89.174417}. Through the definition of $M_L^\perp$ and $H_L^\perp$, 
one can also determine the real susceptibilities $\chi^{xx}=\chi^{yy}=\chi^\prime$, 
and $\chi^{yx}=-\chi^{xy}=\chi^{\prime\prime}$. Note that the magnetization precession 
presents a response in both field directions $H^x$ and $H^y$. The delay effect is caused by the 
spin relaxation and, in the opposite case, when $\eta=0$, the magnetization instantly 
responds to the field application. Hence, in the driven magnetization precession,
the magnetic susceptibility on the spherical surface presents the same behavior observed in
flat models \cite{jpcm32.305802}. Indeed, provided that we are dealing with uniform fields, it is expected
only the $q=0$ (for flat geometry) or $L=0$ (for spherical geometry) susceptibility term. The uniform nature is
then similar in both spaces. On the other hand, if the gradient of the magnetic field is included, the geometry 
influence is evidenced by the non-uniform terms in the expansion of $\chi$.

\section{Summary and conclusions}

In this article, we investigated the thermodynamics of the ferromagnetic model on the spherical surface. 
Similar problems involving flat surfaces have been studied in recent years; however there is a lack of
information about spintronics of ferromagnetic models on curved spaces, and the present work was developed 
for clarifying some points of the theme. 

The Hamiltonian was described by using the SBMFT, which represents the spin operators 
in terms of two kinds of bosons, $a$ and $b$. Provided that we are dealing with the coherent state of
an FM model, we expected a minor influence of fluctuations in the mean-field parameters, and Gaussian
corrections were not included in this work. Opposite to the flat surfaces, we found a finite critical 
temperature $T_c$ that separates the ordered phase from the disordered one. The critical temperature depends
on the density of sites $\sigma$, and we recovery the result $T_c=0$ in the limit $\sigma\to\infty$, which
is the local representation of the two-dimensional flat surface. We showed that, at the low-temperature limit, 
the developed equations give trustworthy results. In addition, we analyzed the magnetization dynamics in the 
presence of two orthogonal magnetic fields; the static field $B^z$ that aligns the spin field and the oscillating
field $B^x(t)$ responsible for the precession motion. We demonstrated systematically that in the presence
of the cited magnetic fields, the model is described by $SU(2)$ coherent states, which are suitable for evaluating
all thermodynamic properties. The magnetization behavior and the magnetic susceptibility were 
evaluated through the coherent states, and the results are in agreement with the expected ones obtained from
the Linear Spin-Wave theory devevolped in Appendix (\ref{appendix_LSW}). 
Indeed, we found that the more efficient magnetization precession occurs at the resonant condition 
$\omega_\textrm{rf}=g\mu_0 H^z$, for instance.

In summary, we showed that the description of ferromagnetism and spintronics on the spherical surface can
be developed through the $SU(2)$ coherent states. Here, we applied the formalism for describing the magnetization
precession; however, other spin experiments on curved space such as spin superfluidity, spin current injection, 
spin-transfer torque, and spin-wave dissipation processes, can also be explained by the developed method.
In addition, magnetic models on the spherical surface could be used for testing general curvature effects in favor
of the complicated experiments on curved space, like the microgravity experiments for verifying BEC on spherical manifold, for example.

\appendix
\section{Path integral for $SU(2)$ coherent states}
\label{appendix_su2cs}
We can use the $SU(2)$ coherent states for evaluating the partition function through the
path integral formalism. Let us adopt the following Hamiltonian
\begin{equation}
H_0=\sum_{\langle ij\rangle}(\epsilon_{ij}^{(a)} a_i^\dagger a_j +\epsilon_{ij}^{(b)} b_i^\dagger b_j),
\end{equation}
written in terms of the Schwinger bosons $a$ and $b$. 

Following the standard path integral procedures, we divide the time interval
in $N$ small $\Delta\tau$ steps, with $\tau_0=0$ and $\tau_N=\beta\hbar$, 
which provide the partition function
\begin{equation}
Z_0=\sum_\zeta\langle\zeta(\tau_N)|\prod_{p=1}^N e^{-\hat{H}_0(\tau_p)\Delta\tau}|\zeta(\tau_0)\rangle,
\end{equation}
with the periodic condition $|\zeta(\tau_N)\rangle=|\zeta(\tau_0)\rangle=\prod_i|\zeta_i\rangle$ 
(here, the $2S$ subscript was omitted for simplifying the notation). Inserting the identity 
$I=\int\ud\zeta|\zeta\rangle\langle\zeta|$ between the intervals, we obtain

\begin{IEEEeqnarray}{L}
\langle\zeta(\tau_p)|e^{-\hat{H}\Delta\tau}|\zeta(\tau_p-\Delta\tau)\rangle\simeq 1-
\langle\zeta(\tau_p)|\dot{\zeta}(\tau_p)\rangle\Delta\tau-\nonumber\\
-H_0(\tau_p),
\end{IEEEeqnarray}
where $H_0(\tau_p)=\langle\zeta(\tau_p)|\hat{H}_0|\zeta(\tau_p)\rangle$ 
is an ordinary real function obtained through the replacement of the
operators $a_i$ and $b_i$ by the respective fields $a_i=\sqrt{2S}u_i$ and 
$b_i=\sqrt{2S}v_i$ [see Eq.(\ref{eq.azeta}) and (\ref{eq.bzeta})]. Note that, although
$|\zeta\rangle$ is not an annihilation eigenstate, as occurs in the $U(1)$ coherent state,
yet is possible for evaluating the average $\langle \exp(-\hat{H}_0\Delta\tau)\rangle$.
The second term in the above equation is the Berry phase 
$\Omega_B=-iS\sum_i[\dot{\varphi}_i(1-\cos\theta_i)]$. It is a straightforward evaluation for
showing $\sum_i(\bar{a}_i\dot{a}_i+\bar{b}_i\dot{b}_i)=\Omega_B$. 
Therefore, the partition function is written as the path
integral $Z_0=\int D[\bar{a},a]D[\bar{b},b] \exp(-\mathcal{S}/\hbar)$, 
where the action is given by

\begin{equation}
\mathcal{S}=\int_0^{\beta\hbar}\ud\tau\left[\hbar\sum_i(\bar{a}_i\partial_\tau a_i+\bar{b}_i\partial_\tau b_i)+H_0(\tau)\right].
\end{equation}

\section{Linear Spin-Wave Approximation}
\label{appendix_LSW}
In order to compare the results of the SU(2) coherent states obtained from the SBMFT, we develop the Linear Spin-Wave theory 
through the Holstein-Primakoff formalism\cite{pr58.1098}. Applying the gauge fixing condition $a=a^\dagger$ in the Schwinger formalism,
one can obtain the HP bosonic representation of the spin operators: $S_i^+=\sqrt{2S-b_i^\dagger b_i}b_i$, 
$S_i^-=b_i^\dagger\sqrt{2S-b_i^\dagger b_i}$ and $S_i^z=S-b_i^\dagger b_i$\cite{jpamg27.2915}. 
At low temperatures, the LSW approximation is expected to provide reasonable
results, and we adopt the approximation $S_i^+\approx \sqrt{2S}b_i$ and $S_i^-\approx\sqrt{2S}b_i^\dagger$. Therefore, the 
quadratic Hamiltonian, given by Eq. (\ref{eq.H0}), is written as
\begin{equation}
H_0=JS\sum_{\langle ij\rangle}b_i^\dagger(b_i-b_j)-\gamma\hbar B^z\sum_i(S-b_i^\dagger b_i), 
\end{equation}
where $\gamma=g\mu_B/\hbar$ is the gyromagnetic ratio. After expanding $b_j$ around $b_i$, the continuum limit is taken into account, which provides the Hamiltonian
\begin{equation}
H_0=\int\ud\Omega b^\dagger\left(-\frac{zJSR^2\nabla^2}{4}+\gamma\hbar\sigma B^z\right)b.
\end{equation}
The spherical harmonic expansion
\begin{equation}
b(\theta,\varphi)=\frac{1}{\sqrt{\sigma}}\sum_L b_L Y_L(\theta,\varphi),
\end{equation}
then results in $H_0=\sum_L (\epsilon_L+\gamma\hbar B^z) b_L^\dagger b_L$, where $\epsilon_L=zJSl(l+1)/4\sigma$ ($L$ is the compact 
representation of $m=-l,-l+1,\ldots,l-1,l$ and $l=0,1,2,\ldots$). This is the 
same spectrum energy obtained from the Schwinger bosonic formalism, see Eq.(\ref{eq.hamiltonian_diag}), in the very low-temperature limit. 
In this limit, we can approximate the mean-field parameter as $F\approx 2S$ and 
adopted the chemical potential $\mu=-\gamma\hbar B^z/2$. Therefore, the a-spinons condensate while spin-wave 
excitations are mapped by the b-spinons of the SBMFT. Applying the same procedure in the interaction (\ref{eq.interaction}), we get
\begin{equation}
V(t)=-\frac{\sqrt{2S\sigma}}{2}\gamma\hbar\sum_L [B_L^x(t) b_L^\dagger+\bar{B}_L^x(t) b_L],
\end{equation}
where $B_L^x(t)$ are the spherical harmonic components of the oscillating magnetic field. In the interaction picture,
the average of the operator $A(t)$ is expressed as $\langle A(t)\rangle=\langle S^\dagger(t)\hat{A}(t)S(t)\rangle$,
where $S(t)=T_t\exp(-i\int \hat{V}(t^\prime)\ud t^\prime/\hbar)$, and the caret denotes time evolution according $H_0$. 
It is a straightforward procedure to demonstrate that $S(t)=\exp[i\Phi+\sum_L \beta_L(t) b_L^\dagger-\bar{\beta}_L(t) b_L]$,
where $\Phi$ is an irrelavant phase and the coherent eigenvalue, which is defined by $b|\beta\rangle=\beta|\beta\rangle$, is given by
\begin{equation}
\beta_L(t)=\sqrt{2\pi S\sigma}\gamma B^x \delta_{L0}\frac{e^{i(\gamma B^z-\omega_\textrm{rf}-i\eta)t}}{\gamma B^z-\omega_\textrm{rf}-i\eta}.
\end{equation}
Here, we consider a monochromatic uniform magnetic field $B^x(t)=B^x e^{-i\omega_\textrm{rf}t}$, and
$\eta$ (the damping factor) is included for ensuring the convergence in the limit $t\to -\infty$. The number of particles 
$N_i=|\beta_i|^2$ is then 
\begin{equation}
N_i=\sum_{L^\prime L}\frac{\bar{\beta}_{L^\prime}\beta_L}{\sigma}\bar{Y}_{L^\prime}Y_L=\frac{(\gamma B^x)^2S}{2[(\gamma B^z-\omega_\textrm{rf})^2+\eta^2]},
\end{equation}
which presents the maximum at the resonanting condition $\gamma B^z=\omega_\textrm{rf}$.
Following the same steps of Sec. (\ref{sec.mag_dynamics}), we evaluate the magnetization dynamics on the spherical surface
using the $U(1)$ coherent states of the HP formalism, which yiedls
\begin{IEEEeqnarray}{rCl}
M_i^\perp(t)&=&\frac{\gamma\hbar}{\varepsilon^2}\langle S_i^+(t)\rangle=\frac{\gamma\hbar}{\varepsilon^2}\sqrt{\frac{2S}{\sigma}}\sum_L \beta_L(t)e^{-i\gamma B^z t}Y_L\nonumber\\
&=&\frac{M_s \gamma B^x(t)}{\gamma B^z-\omega_\textrm{rf}-i\eta},
\end{IEEEeqnarray}
which is identical to Eq. (\ref{eq.magnetization}). The magnetic susceptibility is then determined from the above equation and
we reach the same results obtained from the SBMFT.

\bibliography{manuscript}
\end{document}